\documentclass[dvips,prb,twocolumn,showpacs]{revtex4}
\usepackage{amssymb}
\usepackage[dvips]{graphicx}            

\newcommand{\be}{\begin{equation}}
\newcommand{\ee}{\end{equation}}
\newcommand{\ba}{\begin{eqnarray}}
\newcommand{\ea}{\end{eqnarray}}
\newcommand{\nn}{\nonumber}

\begin{document}

\title{\textbf{ Local Temperatures and Heat Flow in Quantum Driven
    Systems  }}  
\author{Alvaro Caso, Liliana Arrachea and Gustavo S. Lozano}

\address{Departamento de F\'{\i}sica, FCEyN, Universidad de Buenos Aires,
Pabell\'on 1, Ciudad Universitaria, 1428, Buenos Aires, Argentina.}

\begin{abstract}
We discuss the concept of local temperature for quantum systems driven 
out of equilibrium by ac pumps showing explicitly that it is the
correct indicator for heat flow.  
We also show that its use allows for a generalization of the
Wiedemann-Franz law. 
\end{abstract}

\maketitle

\section{Introduction}

In the last years growing research activity has focused on the search
for a better understanding of the mechanisms for heat production and
energy flow in non equilibrium quantum systems at the microscopic
level.   
Examples are thermoelectric effects in quantum point contacts,
\cite{thermoqpoint} quantum pumps under driving induced with ac
voltages acting at the walls, \cite{liliheatpump} ``quantum''
capacitors, \cite{heatquantumcap} driven small-size heterostructures,
\cite{pekkola} as well as atomic and molecular junctions
\cite{dubi-diventra}, nanomechanical systems, \cite{nanomec} and
photonic systems. \cite{phot} 
The understanding of the entropy production and its connection with
the non equilibrium dynamics has also been a central subject of
research  in other areas of physics, including aging regimes in glassy
systems, sheared glasses, granular materials, and colloids.
\cite{fdr,cukupel,letoandco} 
A very successful concept in the characterization of non-equilibrium
states concerns the definition of an ``effective temperature.''
In glassy systems the definition of an effective temperature was
introduced via generalized fluctuation-dissipation relations
\cite{fdr} and the validity of such a temperature as a physical
meaningful concept was further supported by showing that such a
temperature coincides with the one that the measurement with a
thermometer casts for that system.\cite{cukupel} 

The definition of an effective temperature from a
fluctuation-dissipation relation in quantum models was introduced in
Ref.~\onlinecite{letogus} for glassy systems and later explored for
electronic systems in Ref.~\onlinecite{lilileto}.  
In this last work a ring threaded by a linear-in-time-dependent
magnetic flux in contact to a reservoir was studied. 
The underlying physics is the induction of a constant electromotive
force and generation of a current with a dc component, with the
concomitant heat dissipation into the reservoir by the Joule effect. 
On the basis of a numerical analysis, it was found that the so defined
effective temperature of the driven ring was larger than that of the
reservoir, in consistency with the idea that the driving heats the
ring and the energy is dissipated toward the reservoir.

In a recent work\cite{cal} we have addressed the issue of
identifying effective temperatures in the context of transport in
electronic quantum systems driven out of equilibrium by external
(periodic) pumping potentials. 
Examples of this type of system are quantum dots with ac voltages
acting at the walls (quantum pumps) \cite{pump} and quantum
capacitors,\cite{qcapexp} which display energy transport regimes much
richer than the case of the ring described above. 
In fact, these systems can not only dissipate energy in the form of
heat but can also pump energy between the different reservoirs,
generating refrigeration.  
We have defined a ``local'' temperature along these set-ups by
introducing a thermometer, i.e., a macroscopic system which is in
local equilibrium with the system, even when the system itself is out
of equilibrium. 
This is the thermal analog of the voltage probe discussed in
Refs. \onlinecite{fourpoint} and \onlinecite{fourpointfed}.  
On the other hand we have also defined an effective temperature by
analyzing a local fluctuation-dissipation relation.  
Interestingly enough, we have been able to show that the two
definitions of the temperature coincide when the ac driving is weak,
i.e., for low amplitude and frequency of the ac voltages.  
The behavior of the local temperature along the setup is also very
interesting on its own.  
It displays oscillations modulated by $2 k_F$, $k_F$ being the Fermi
vector. 
This feature has been also observed in the behavior of the local
temperature in systems under stationary transport
\cite{dubi-diventra} and must be interpreted as a signature of the
coherent nature of the electronic transport along the structure, where
scattering processes with the ac potentials generate an interference
pattern.  
It is the counterpart in the framework of the energy propagation to
the Friedel oscillations detected when the structure is sensed with a
local voltage probe. \cite{fourpoint, fourpointfed}
Remarkably, in some situations, it is possible to distinguish regions
of the structure with a local temperature that is  cooler than that of
the reservoirs.   

The aim of the present work is to further investigate the scope
of the concepts of local and effective temperature in quantum driven 
systems. 
In particular, our goal is to show that such a parameter verifies the
thermodynamical properties of a temperature, in the sense that it
signals the direction for heat flow. 
We also  slightly generalize the definition of the thermometer, by
allowing it to act simultaneously as a thermal probe and a voltage
probe.   
Finally, we show that the effective temperature plays a fundamental
role in a generalization of the Wiedemann-Franz law to an out of
equilibrium set-up.  

The work is organized as follows. 
In Sec.~II, we present the model and summarize the theoretical
treatment.
In Sec.~III we present results. 
In Sec.~IV  we generalize the model for the thermometer. 
Section V is devoted to discussion and conclusions. We give some
details of the calculation in the Appendix.

\section{Model and theoretical treatment}

\begin{figure}
\centering
\includegraphics[width=80mm,angle=0,clip]{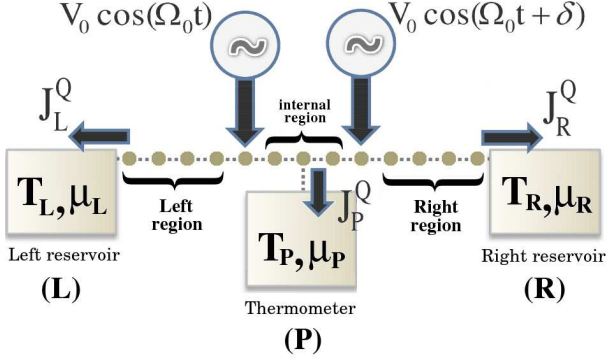} {\small {} }
\caption{{ Scheme of the set up. The central device is a wire
    connected to the left and right. The third reservoir ($P$) 
    represents the thermometer, which consists of a macroscopic system
    weakly coupled to a given point of the central device.}}
\label{fig0}
\end{figure}

We consider here the same set-up  as in
Ref.~\onlinecite{cal} which we display in Fig.~\ref{fig0}.
This is a quantum  driven system described by a Hamiltonian
$H_{sys}(t)$ and  a thermometer characterized by a Hamiltonian $
H_{P}$ that are {\it locally} coupled via $H_{cP}$ in such a way that
the total Hamiltonian can be written as
\be
H(t) = H_{sys}(t)+H_{cP}+H_{P}.
\ee
For the driven system we take a device composed of a central
part [$H_C(t)$] and two reservoirs ($ H_{L}, H_{R}$), coupled to the
central part via contacts ($ H_{cL}, H_{cR}$), 
\be
H_{sys}(t) = H_{L}+H_{cL}+ H_{C}(t)+H_{cR}+ H_{R}.
\ee
The  Hamiltonian describing the central system ($C$) contains  the ac
driving fields, $H_C(t)=H_0+ H_V(t)$. 
We assume that $H_0$ is a Hamiltonian for non interacting electrons
while $H_V(t)$ is harmonically time dependent, i.e.,
$H_V(t)= \sum_{n=-\infty}^{+\infty} e^{-i \Omega_0 n t} H_V(n)$. We
leave further details of the model for the moment undetermined in
order to make the coming discussion as model independent as possible. 

Both reservoirs and the local probe are modeled by systems of
non interacting electrons with many degrees of freedom: $ 
H_{\alpha}=
\sum_{k \alpha } \varepsilon_{k \alpha} c^{\dagger}_{k \alpha} c_{k
\alpha}$, where $\alpha=L,R,P$. The corresponding contacts are $H_{c
\alpha}= w_{c \alpha} (c^{\dagger}_{k \alpha} c_{l
\alpha}+c^{\dagger}_{l \alpha} c_{k \alpha})$, where $l \alpha$
denotes the coordinate of $C$ at which the reservoir $\alpha$ is
connected. 
As in previous works,\cite{cal,fourpoint,fourpointfed}  we consider a
non invasive probe, which implies that $w_{cP}$ is small enough to be
treated at the lowest order of perturbation theory when necessary. 

To describe the dynamics of the system we use the
Schwinger-Keldysh-Green functions formalism.   
This involves  the calculation of the Keldysh and retarded Green
functions, 
\begin{eqnarray}
G^{K}_{l,l'}(t,t')&=& i \langle c^{\dagger}_{l'}(t') c_l(t) - c_l(t)
c^{\dagger}_{l'}(t')  \rangle, 
\nonumber \\
G^R_{l,l'}(t,t')&=& -i \Theta(t-t') \langle c_l(t)
c^{\dagger}_{l'}(t')+c^{\dagger}_{l'}(t') c_l(t) 
\rangle ,
\label{green}
\end{eqnarray}
where the indexes $l,l'$ denote spatial coordinates of the central
system. 
These Green functions can be evaluated after solving the Dyson
equations. 
For the case of harmonic driving it is convenient to use the
Floquet-Fourier representation of the Green functions: \cite{liliflo}
\begin{equation}
 G_{l,l'}^{K,R}(t, t-\tau) = \sum_{k=-\infty}^{\infty}
 \int_{-\infty}^{\infty} \frac{d\omega}{2 \pi} 
 e^{-i (k \Omega_0 t + \omega \tau)}  G_{l,l'}^{K,R}(k,\omega).
\end{equation}

\section{Defining the temperature}
\subsection{Local temperature determined by a thermometer}
Heat transport through the central system can occur due to a
temperature or chemical potential difference between the  reservoirs
as well as as the result of pumping by the external sources.  
In a generic situation, if the probe is connected to the central
system, there is also heat exchange between the system and the
probe. 
In Ref.~\onlinecite{cal} the local temperature $T_{lP}$ was defined as 
the value of $T_P$ (i.e., the temperature of the probe) such that heat 
exchange between the central system and the probe vanishes. 

It can be  shown\cite{liliheatpump} that, given $H_C(t)$ without
many-body interactions, the heat current from the central system and
the thermometer can be expressed as ($\hbar=k_B=e=1$) 
\ba
& &J_P^Q = \sum_{\alpha=L,R,P} \sum_{k=-\infty}^{\infty}
\int_{-\infty}^{\infty} \frac{d\omega}{2 \pi} 
 \{  [f_\alpha(\omega)-f_P(\omega_k)] \nonumber \\
& &   \times (\omega_k - \mu) \Gamma_P(\omega_k) \Gamma_\alpha(\omega)
 \left| G^R_{lP,l\alpha}(k,\omega)\right|^2 \}, \label{jq}
\ea
where $\omega_k=\omega+k\Omega_0$ and $\Gamma_{\alpha}(\omega) = -2
\pi |w_{\alpha}|^2 \sum_{k \alpha} \delta(\omega-\varepsilon_{k
  \alpha})$ are the spectral functions that determine the escape to 
the reservoirs ($\alpha = L, R, P$), and $f_\alpha(\omega)=
1/[e^{\beta_{\alpha}(\omega -\mu_{\alpha})}+1]$ is the Fermi
function, which depends on $T_{\alpha}=1/\beta_{\alpha}$ and
$\mu_\alpha $ the temperature and the chemical potential  of the
reservoir $\alpha$.  
 Thus, the local temperature $T_{lP}$ corresponds to the solution of
 the equation 
\begin{equation}\label{tlp}
J_P^Q(T_{lP}) = 0. 
\end{equation}

In Ref.~\onlinecite{cal} the value of $\mu_P$ was kept fixed (and
equal to that of the reservoirs, $\mu_L=\mu_R=\mu_P$).   
Our thermometer, however, is a reservoir not only for energy but also
for particles. 
In fact, the same setup but with the role of temperature and chemical
potential exchanged was considered in
Refs.~\onlinecite{fourpoint} and \onlinecite{fourpointfed} to define
the local voltage of a driven structure.    
One question that arises is how the situation gets modified when we
allow {\em both} the temperature {\em and} the voltage of the probe to 
adjust simultaneously to define the local temperature and the local
voltage.  
Such a procedure has been followed in Ref. \onlinecite{polianski}.  
Thus, in an analogous way as we did before, we now define the local
temperature $T_{lP}^*$ (where we use the $*$ symbol to distinguish it
from the definition above) and local voltage $\mu_{lP}^*$,
respectively, as the temperature and the voltage of the probe that
vanish simultaneously both the charge and the heat currents between
the system and the probe, i.e.,  
\be
\left\{ \begin{array}{rcl}
J^Q_P (T_P^*,\mu_P^*) & = & 0, \\\nonumber
J^e_P (T_P^*,\mu_P^*) & = & 0,
\end{array} \right.
\label{tstar}
\ee
where (see Refs. \onlinecite{liliflo,fourpointfed})
\ba
& &J_P^e = \sum_{\alpha=L,R,P} \sum_{k=-\infty}^{\infty}
\int_{-\infty}^{\infty} \frac{d\omega}{2 \pi} 
 \{  [f_\alpha(\omega)-f_P(\omega_k)] \nonumber \\
& &   \times 
\Gamma_P(\omega_k) \Gamma_\alpha(\omega) \left|
G^R_{lP,l\alpha}(k,\omega)\right|^2 \}. \label{je} 
\ea

\subsection{Effective temperature from a Fluctuation-Dissipation
  Relation} 
For systems in equilibrium, the fluctuation dissipation theorem
establishes a relation between the Keldysh (correlation) and retarded
Green functions. 
Indeed, for a system like the one under consideration, but without the
time-dependent fields, it can be shown that  the relation between the
fluctuations in the system,  $iG_{l,l}^{0,K}(\omega)$, with the
dissipation term of the bath, $\Gamma_{\alpha}(\omega)$, is
\cite{letogus,lilileto}
\ba \!\!\!\!\! & &iG_{l,l}^{0,K}(\omega)=
\tanh[\frac{\beta(\omega-\mu)}{2}] \varphi^0_l(\omega), \label{fdt1}
\\
\!\!\!\!\!\!
& & \varphi^0_l(\omega)=-2
\mbox{Im}[G_{l,l}^{0,R}(\omega)]= \sum_{\alpha=L,R} |G^{0,R}_{l,l
\alpha}(\omega)|^2 \Gamma_{\alpha}(\omega),
 \ea
where the index $^{0}$ indicates that we are considering the
equilibrium  system, i.e. with the term $H_V(t)=0$ and all the  
reservoirs at the same temperature $T=1/\beta$. 

In the presence of time-dependent voltages it can be shown that
\ba 
iG^K_{l,l}(0,\omega)&=& \sum_{k = -\infty}^{\infty} \tanh
[\frac{\beta(\omega_{-k} - \mu)}{2} ] 
\varphi_l(k,\omega_{-k}),\label{fdtg2} \\ 
 \varphi_{l}(k,\omega) & = &  \sum_{\alpha= L,R}
 \left|G^R_{l, l \alpha}(k,\omega)\right|^2 \Gamma_{\alpha}(\omega).
\ea 

In Ref.  \onlinecite{cal} we have shown  that within the weak
driving-adiabatic regime, where the term $H_V(t)$ is treated as a
perturbation and the driving frequency is smaller than the dwell time
of the electrons within the central system \cite{adia}, it is possible
to define an effective temperature $T^{eff}_l=1/\beta^{eff}_l$ through
the following relation: 
\ba 
iG^K_{l,l}(0,\omega)-iG^K_{l,l}(0,\mu)& = &
\tanh[\frac{\beta^{eff}_l (\omega-\mu)}{2}] \overline{\varphi_{l}}(
\omega), \label{fdr} 
\ea 
with $\overline{\varphi_{l}}( \omega)=-2 \mbox{Im}
[G_{l,l}^R(0,\omega)]= \sum_k \varphi_l(k,\omega_{-k})$. 
A similar relation in the time domain has been studied numerically for
a driven ring in contact with a reservoir. \cite{lilileto} 

\section{Results}
In this section we present results for a central device consisting of
non-interacting electrons in a one-dimensional lattice:
\be
H_0= -w \sum_{l,l^{\prime}} (c^\dagger_{l} c_{l^{\prime}} + h. c.),
\ee
where $w$ denotes a hopping matrix element between neighboring
positions $l,l^{\prime}$ on the lattice, and a driving term of the
form: 
\be \label{hv}
 H_V(t)= \sum_{j=1}^2 e V_j(t) c^\dagger_{lj} c_{lj} ,
\ee
with $V_j(t)= V_0 \cos( \Omega_0 t + \delta_j)$, $lj$ being the
positions at where two ac fields oscillating with the same frequency 
and a phase-lag are applied. 
This defines a simple model for a quantum pump where two ac gate
voltages are applied at the walls of a quantum
dot. \cite{liliflo,adia,pump}  

\subsection{Equivalence between the different definitions of the
  temperature at weak driving} 
In Ref. \onlinecite{cal} we have analyzed the weak driving, which
corresponds to the ac voltage amplitudes lower than the kinetic energy
of the electrons in the structure and the driving frequency lower than
the inverse of the dwell time of these electrons. 
We have analytically shown in this case that the local temperature
defined from Eq. (\ref{tlp}), with the chemical potential of the
reservoir kept fixed, is identical to the effective temperature
defined from the local fluctuation-dissipation relation given by
Eq. (\ref{fdr}).  
That is,  
\begin{equation}
T^{eff}_{lP}=T_{lP}.
\end{equation}  

In Sec. \ref{def1} of the Appendix we summarize the main steps leading to this
result and we also show that within the weak-driving regime the local
temperature can be expressed as
\ba
T_{lP}^2 &  & \sim T^2+ \frac{3}{\pi^2} \lambda_{lP}^{(0)}(\mu)
\Omega_0^2 +  2 \lambda_{lP}^{(1)}(\mu) T^2 \Omega_0 
\nonumber \\
& & - \frac{1}{2} \lambda_{lP}^{(2)}(\mu) T^2 \Omega_0^2,
 \label{tloc}
\ea
with
\be \label{lambda}
\lambda_{l}^{(n)}(\omega)=  \frac{1}{ \sum_{k=-1}^{1}
  \varphi_{l}(k,\omega)} 
\sum_{k=-1}^{1} (k)^{n+2} \frac{d^n[ \varphi_{l}(k,\omega)]}{d
  \omega^n} , 
\ee
\be 
\varphi_{l}(k,
\omega)  =   \sum_{\alpha= L,R}
 \left|G^R_{l, l \alpha}(k,\omega)\right|^2 \Gamma_{\alpha}(\omega).
\ee 

By keeping only the lowest order in $\Omega_0$, the local temperature
$T_{lP}$ can be cast into the form   
\be
T_{lP} = T \left[ 1 + \lambda^{(1)}_{lP}(\mu) \Omega_0
  \right]. \label{T0} 
\ee
    
Analytical expressions for $T_{lP}^*$ defined in Eqs.  (\ref{tstar})
are considerable harder to obtain than those $T_{lP}$. 
Nevertheless, we have been able to show that within the regime of
interest (see Sec. \ref{def2} of the Appendix for details)

\be
T_{lP}^* = T \left[ 1 + \Omega_0 \frac{d}{d\omega}
\left( \frac{\sum_{k=-1}^1 k \varphi_{lP}(k,\omega)}
{\sum_{k=-1}^1 \varphi_{lP}(k,\omega)}
\right)_{\omega=\mu}\right].\label{T1} 
\ee
It is easy to see that for $\left. \frac{d}{d\omega}
\Gamma_\alpha \right|_{\omega = \mu} \sim 0$, which is in general
satisfied for metallic electrodes with a featureless band,
Eq. (\ref{T1}) becomes 
\be \label{def2DeltaT}
T_{lP}^* = T \left[ 1 + \lambda^{(1)}_{lP}(\mu) \Omega_0
  \right]=T_{lP}.
\ee

Thus,  it is possible to prove that all the three definitions of the
local temperature, $T_{eff}$ from a fluctuation-dissipation relation,
$T_{lP}$ from a thermometer, and $T^*_{lP}$ from a thermometer that is
also a voltage probe, coincide within the weak-driving regime.

\subsection{ Temperature and the direction for heat flow}  

We now turn  to explore the relation between the  local temperature
and heat flow between the central system and the left and right
reservoirs.    

\begin{figure}
\centering
\includegraphics[width=80mm,angle=0,clip]{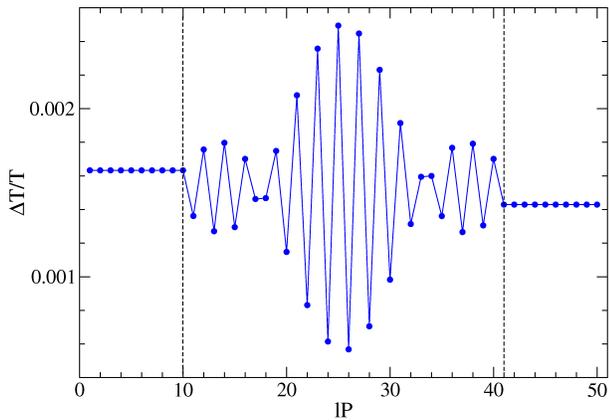} {\small {} }
\caption{{
(Color online) Local temperature along a one-dimensional model of
    $N=50$ sites with two ac fields operating with a phase lag of
    $\delta=\pi/2$ at the positions indicated by dotted lines. The
    system is in contact with reservoirs with chemical potentials
    $\mu=0.2$ and temperature $T=0.025$. The driving frequency is
    $\Omega_0=0.05$ and the amplitude is $V_0 =0.05$. }}
\label{fig1}
\end{figure}

In Fig.~\ref{fig1} we show  a typical temperature profile along the
structure. 
The value of $T_{l}$ is plotted for each point of the chain, for
$T_L=T_R=T$, $\mu_L=\mu_R=\mu$,  and a particular low value
$\Omega_0$. 
We  can distinguish two regions within the central structure, denoted
as ``Left'' and ``Right'' regions in Fig. \ref{fig0}, which are
defined between the contact with the left (right) reservoir and the left
(right) pumping centers. The local temperature at weak driving is
constant within these regions but different from the one of the
reservoirs. In the internal region between the two pumping centers,
the local temperature displays $2k_F$ Friedel-like oscillations,
$k_F$ being the Fermi vector of the electrons leaving the reservoirs. 
This feature is similar to the one observed in other small size
structures under stationary driving \cite{dubi-diventra} and has the
same origin as the oscillations in the local voltage profile sensed by
a voltage probe, \cite{fourpoint,fourpointfed} namely, the
interference generated by elastic scattering processes at the two
pumping centers.   

\begin{figure}
\centering
\includegraphics[width=80mm,angle=0,clip]{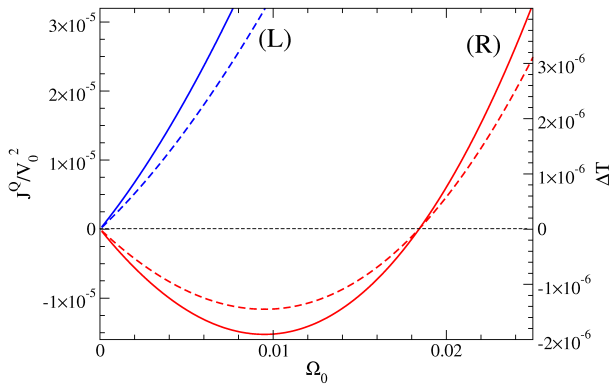} {\small {} }
\caption{{
(Color online) dc heat current divided by $V_0^2$ (solid) and local
    temperature difference (dashed) between the system and the left
    reservoir (blue) or the right reservoir (red) as a function of
    driving frequency $\Omega_0$. The phase lag is $\delta=1.88$ and
    the driving amplitude is $V_0=0.05$. The temperature and the
    chemical potential of the reservoirs are $T=0.025$ and $\mu=0.2$.
 }}
\label{fig3}
\end{figure}

We would like to explore whether heat flow through the  contacts to
the reservoirs is described by a relation of the type
\begin{equation} \label{thercond}
J^Q_\alpha = K_\alpha \Delta T_\alpha,
\end{equation} 
as it happens in systems where the heat flow is induced by an explicit
temperature gradient. 
In our case, the gradient is defined as $\Delta T_{\alpha} =
T_{l\alpha}-T_\alpha$, $T_{l\alpha}$ being the local temperature at
the point of the central device connected to the $\alpha$ reservoir,
while $K_\alpha$ is a {\it positive} effective contact thermal
conductance.   
Thus, we evaluate independently the dc components of the heat currents  
between the system and each of the reservoirs, as well as the
local temperatures at the contacts.
Results for heat flow and local temperature gradients
$\Delta T_\alpha$ are shown in Fig.~\ref{fig3}, as functions of 
the pumping frequency for reservoirs with the same temperature $T$ and
the same chemical potential $\mu$.  
Since the dc heat current is $\propto V_0^2$ for low driving
amplitudes, we found it convenient to show $J^Q/V_0^2$ in the figure. 
The flow is defined as positive (negative) when the heat flows to
(from) the reservoir.

The behavior of the heat flow at the left reservoir ($L$)  corresponds
to  a situation in which heat enters the reservoir.  
This is associated with the idea of heat flowing from a hot region to
a colder one. 
Correspondingly, the local temperature at the contact point of the
system is higher than $T_L$.  

Nevertheless, in a pumping regime, we expect to find situations in
which  heat can be extracted from one reservoir to be pumped  into the
system and the other reservoir. 
This is indeed the situation for the right lead ($R$), where for very
low frequencies the heat flow is negative. 
In the same figure we show that the corresponding gradient of
temperature along the contact shows a behavior compatible with the
heat flow.  
That is, $T_{lR}$ is lower than $T_R$. 

For higher frequencies, the heat flows into the two reservoirs. 
This is the most common situation, where the central system becomes
heated by the driving voltage and the generated heat is dissipated
into the reservoirs. 
In this regime, the behavior of the gradient of temperature along the
contact also exactly follows the direction of the heat flow. 
In particular, notice in the figure that both $J^Q_R$ and $\Delta T_R$
change the sign exactly at the same frequency.     

\begin{figure}
\centering
\begin{tabular}{c}
\includegraphics
[width=80mm,angle=0,clip]{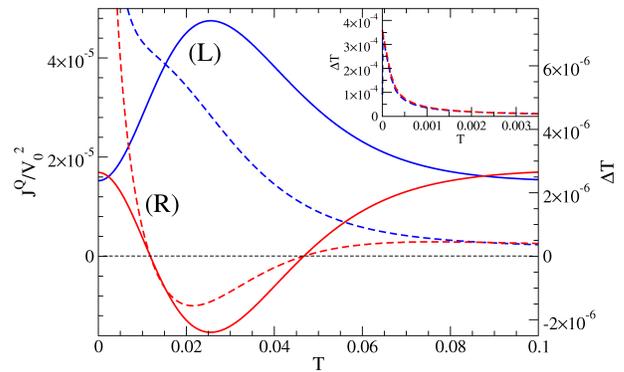} \\\\\\\\
\includegraphics
[width=80mm,angle=0,clip]{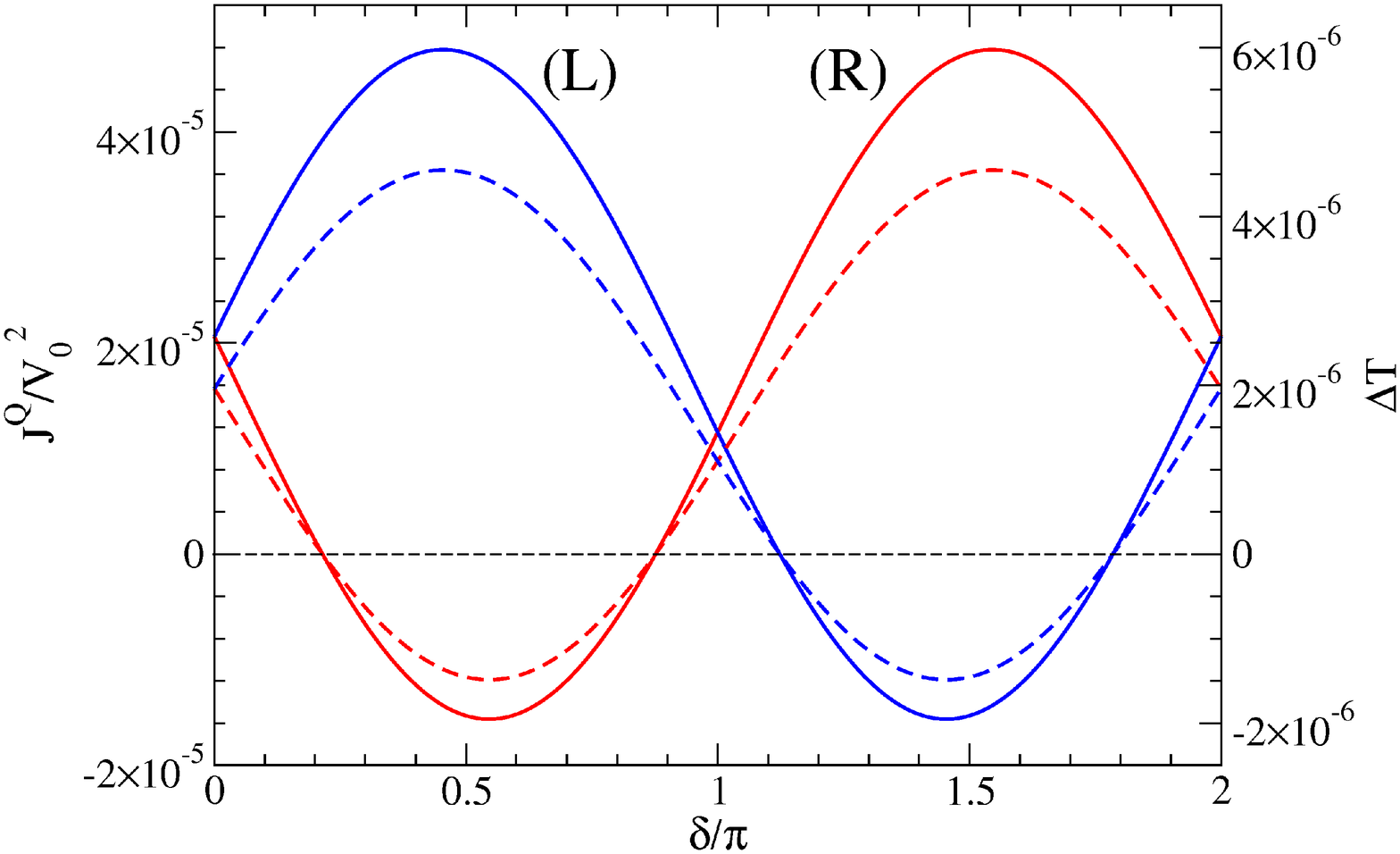}
\end{tabular}
\caption{{
(Color online) Upper panel: Heat flow (solid) and local temperature
    difference (dashed) between the system and the left reservoir
    (blue) or the right reservoir (red) as a function of the
    temperature $T$ of the reservoirs. The phase lag is
    $\delta=\pi/2$, the driving frequency is $\Omega_0=0.01$ and the
    amplitude is $V_0=0.05$. The chemical potential of the reservoirs
    is $\mu=0.2$. 
    Lower panel: Heat flow (solid) and local temperature difference
    (dashed) between the system and the left reservoir (blue) or the
    right reservoir (red) as a function of the phase lag. The driving
    frequency is $\Omega_0=0.01$ and the amplitude is $V_0=0.05$. The
    temperature and the chemical potential of the reservoirs are
    $T=0.025$ and $\mu=0.2$. 
 }}\label{fig4}
\end{figure}

The existence of the pumping regime, requires a delicate interplay
between pumping frequency, temperature, and phase lag but in all cases
we found that the behavior of $\Delta T_{\alpha}$ agrees with that
expected from considerations of heat flow.  
In Fig.~\ref{fig4} we show the heat flow as a function of $T$ and as 
a function of phase lag $\delta$. 
As expected from the symmetries of the set-up, a change of phase
$\delta\rightarrow 2\pi-\delta$ enforces $L\rightarrow R$.  
In all the cases, the behavior of the heat flow is in complete
agreement with Eq. (\ref{thercond}).

\subsection{Generalized Wiedemann-Franz law}

Another interesting property which points toward the identification
of $T_{l\alpha}$ with a {\em bona fide} temperature concerns a
generalization of the Wiedemann-Franz law which we discuss next.  
In addition to the thermal conductance  defined above, we can
consider the voltage probes as in
Refs.~\onlinecite{fourpoint,fourpointfed} to calculate the local 
voltage at the contact and define the effective electrical contact
conductance as follows:  
\be
G_{\alpha}=\frac{J^e_\alpha}{\Delta \mu_{\alpha}},
\ee
where 
\be
\Delta \mu_{\alpha}=\mu_\alpha-\mu_{l{\alpha}},
\ee
where $\mu_{\alpha}$ is the chemical potential of reservoir $\alpha$
and $\mu_{l\alpha}$ is the local chemical potential of the central
system site connected to reservoir $\alpha$. 
As in the previous section we consider $T_{\alpha}=T$ and
$\mu_{\alpha}=\mu$ for $\alpha=L,R$.  

In order to calculate $K_\alpha$ we need the heat current
$J^Q_\alpha$ that flows into the reservoir $\alpha$ and $\Delta 
T_\alpha$. We focus on the weak-driving regime. 
For non invasive thermometers, the heat current that flows into the
reservoir $\alpha$ is 
\be \label{heatA}
J^Q_\alpha = \sum_k \int \frac{d \omega}{2 \pi}
(f(\omega)-f(\omega_k)) (\omega_k - \mu) \tilde{\varphi}_\alpha
(k,\omega), 
\ee
where
\be \label{varphiTilde}
\tilde{\varphi}_{\alpha}(k, \omega) = \sum_{\beta=L,R}
\Gamma_\alpha(\omega_k) \left| G^R_{l\alpha,l\beta}(k,\omega)\right|^2
\Gamma_\beta(\omega).  
\ee

If the temperature $T$ of the reservoirs is small compared to their
Fermi energy, we can apply the Sommerfeld expansion up to order 
$T^2$. 
The low-driving-frequency assumption is introduced by expanding all
the terms of Eq. (\ref{heatA}) in powers of $\Omega_0$. Under these
conditions, the heat current can be rewritten as follows:
\ba
J^Q_\alpha & = & \frac{1}{4\pi} \sum_k \left\{ k^2 \Omega_0^2
\tilde{\varphi}_\alpha(k,\mu)  + T^2 \frac{\pi^2}{3} \left[ 2
  \frac{d \tilde{\varphi}_\alpha}{d\omega}(k,\mu) 
  \right. \right. \nn \\
& & - \left. \left. \frac{1}{2}
  \frac{d^2\tilde{\varphi}_\alpha}{d\omega^2} 
  (k,\mu) k \Omega_0 \right] k \Omega_0 \right\}. 
\label{jq1}
\ea

For high temperature compared to the driving ($T \gg \Omega_0$),
$J^Q_\alpha$ and $\Delta T_\alpha$ [see Eqs. (\ref{def2DeltaT}) and
(\ref{jq1})] are  
\ba
J^Q_\alpha & = & T^2 \frac{\pi}{6} \sum_k k
\frac{d\tilde{\varphi}_\alpha}{d\omega}(k,\mu) \Omega_0, 
\\
\Delta T_\alpha & = & T \lambda^{(1)}_{l\alpha}(\mu)\Omega_0 .
\ea

Using the definitions of $\tilde{\varphi}_\alpha$ and
$\lambda^{(1)}_{l}$ given in Eqs. (\ref{varphiTilde}) and
(\ref{lambda}), respectively, it is easy to show that the thermal
conductance is
\be
K_\alpha = \frac{\pi}{6} \tilde{\varphi}_\alpha(\mu) T,
\label{k1}
\ee
where 
\be
\tilde{\varphi}_\alpha(\mu) = \sum_k \tilde{\varphi}_\alpha(k,\mu).
\ee

The electrical conductance  can be calculated in an analogous
way. Applying the Sommerfeld expansion, expanding all the terms in
powers of $\Omega_0$, and keeping up to first order, the charge current
that flows into reservoir $\alpha$ can be written in the following
way:  
\be
J^e_\alpha = \frac{1}{2 \pi} \sum_k k  
\tilde{\varphi}_\alpha (k,\mu) \Omega_0 .
\ee

For $\Delta \mu_\alpha$, the expression is [see Eq. (\ref{def2T0})]
\be
\Delta \mu_\alpha = \frac{\sum_k k \varphi_{l\alpha}(k,\mu)}{\sum_k 
  \varphi_{l\alpha}(k,\mu)} \Omega_0 .
\ee
Hence, the electrical conductance $G_\alpha$ is
\be
G_\alpha = \frac{1}{2\pi} \tilde{\varphi}_{\alpha}(\mu).  
\label{g1}
\ee
It is possible to show that this result for the electrical conductance
is actually valid for all temperatures. 

At this point it may be convenient to restate units in order to make
it easier to extract useful information from this result. 
Then, from Eqs. (\ref{k1}) and (\ref{g1}) it follows that for the
weak-driving regime, where $T \gg \Omega_0 $, the thermal and
electrical conductances satisfy the Wiedemann-Franz law 
\be
\frac{K_{\alpha}}{G_{\alpha}}= \frac{\pi^2}{3} \left( \frac{k_B}{e}
\right)^2 T .
\ee
\begin{figure}
\centering
\includegraphics[width=80mm,angle=0,clip]{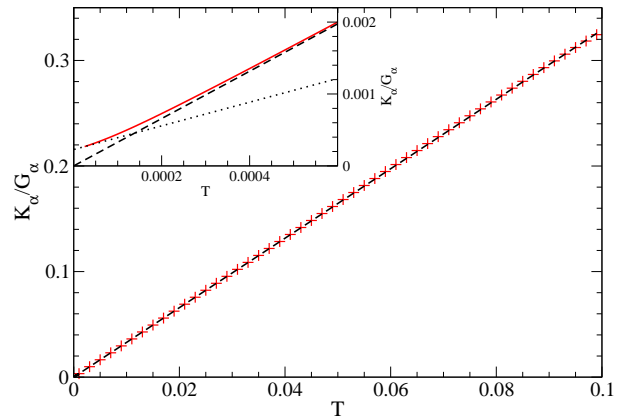}
\caption{{
(Color online) $K/G$ for the right contact (red crosses) as a function
    of temperature.  
    The black dashed line represents the Wiedemann-Franz law.
    Inset: The black dotted line represents the behavior of the
    quotient $K/G$ (red line) for very low temperatures, as depicted
    in Eq. (\ref{K/Gzero}).  
    The phase-lag is $\delta=\pi/2$, the driving frequency is
    $\Omega_0=0.005$ and the amplitude is $V_0=0.05$. 
    The chemical potential of the reservoirs is $\mu=0.2$.
}}
\label{fig6}
\end{figure}
In Fig.~\ref{fig6} we show the ratio $K_\alpha/G_\alpha$ for $\alpha=R$
as a function of temperature $T$. 
The curve for the left reservoir is identical and it is not shown.
We see that for very low $T$, the Wiedemann-Franz law is not
satisfied.
In the low-temperature regime where $ T \ll \Omega_0$, from
Eqs. (\ref{jq1}) and (\ref{def2T0}) it follows that $J^Q_\alpha$ and 
$\Delta T_\alpha$ can be written as 
\ba
J^Q_\alpha & = & \frac{1}{4\pi} \sum_k k^2 \Omega_0^2
\tilde{\varphi}_\alpha(k,\mu), 
\\
\Delta T_\alpha & =  & \sqrt{\frac{3}{\pi^2}
  \lambda^{(0)}_{l\alpha}(\mu)} \Omega_0 - T. 
\ea
Hence, the effective thermal conductance $K_\alpha$ is
\be \label{Kzero}
K_\alpha = \frac{1}{4 \sqrt{3}} \frac{\sum_k k^2 
\tilde{\varphi}_\alpha(k,\mu)}{\sqrt{\lambda^{(0)}_{l\alpha}(\mu)}}
\Omega_0 + \frac{\pi}{12} \tilde{\varphi}_\alpha(\mu) T.
\ee
In this equation it is important to remark that  the thermal
conductance is finite even when the temperature $T$ of the reservoirs
equals zero.

From Eqs. (\ref{g1}) and (\ref{Kzero}) it follows that for low
temperature the quotient $K_\alpha/G_\alpha$, to the lowest order in
$\Omega_0$ and $T$, is  
\be \label{K/Gzero}
\frac{K_{\alpha}}{G_{\alpha}}= \frac{\pi}{2 \sqrt{3}}
\sqrt{\tilde{\lambda}^{(0)}_\alpha(\mu)}\ \Omega_0 + 
\frac{\pi^2}{6} T ,
\ee
where
\be
\tilde{\lambda}^{(0)}_\alpha(\omega) = \frac{1}{\sum_k 
  \tilde{\varphi}_\alpha(k,\omega)}{\sum_k k^2 \tilde{\varphi}_\alpha
  (k,\omega)}. 
\ee
Using the value of $\Delta T_\alpha$ for $T=0$ given in
Eq. (\ref{def2T0}) we can rewrite Eq. (\ref{K/Gzero}), with units
restated as
\be \label{K/Gzero2}
\frac{K_{\alpha}}{G_{\alpha}}= \frac{\pi^2}{6} \left( \frac{k_B}{e}
\right)^2 \left[ \Delta T_\alpha |_{T=0} + T \right].
\ee

From this equation we can see that as the temperature $T$ of the
reservoirs goes to zero, the quotient $K_\alpha/G_\alpha$ approaches
linearly a finite value, explaining the behavior observed in
Fig.~\ref{fig6}.

\section{Summary and Conclusions}

In this work we have analyzed the relation between different
definitions of temperature in a nonequilibrium setup and its
physical meaning, mainly in connection with heat flow. 
More specifically, we have generalized the definition of local
temperature introduced in Ref. \onlinecite{cal} to allow for the
thermometer to act also as a voltage probe and we have shown that in
the situation of interest, i.e, weak driving (small deviations from
equilibrium) and weak system-thermometer coupling (i.e., noninvasive
probe), both definitions coincide, and consequently, both definitions
give the same value as the effective temperature introduced by the
fluctuation-dissipation relation. 

We have also shown that within the low-driving regime, it is possible
to define an effective contact thermal conductance as the quotient
between the dc heat current flowing through a given contact to a
reservoir  and the effective temperature gradient  defined as the
difference between the local temperature at the contact point of the
system and the temperature of the reservoir. 
The behavior of such an effective temperature gradient exactly follows
the direction of the heat flow between the system and the reservoirs.
This is consistent with the idea that the local temperature at the
contact does behave as a  {\em bona fide} temperature.


\begin{acknowledgments}
We acknowledge support from CONICET, ANCyT, UBACYT, Argentina and
J. S. Guggenheim Memorial Foundation (LA). 
\end{acknowledgments}

\appendix*

\section{Analytical expression for local temperature}\label{deducT}

\subsection{Local temperature determined with fixed chemical
  potential of the thermometer} \label{def1}

In this section we present the detailed calculation of the local
temperature, within the adiabatic, low-temperature, and weak-driving
regimes. 

In the weak-driving regime we only keep the terms up to order
$(V_0)^2$ (i.e., Floquet-Fourier components with $k=-1,0,1$). 
Treating the coupling to the thermometer $w_{cP}$ at the lowest order
in perturbation theory and considering that the spectral function of
the thermometer $\Gamma_P(\omega)$ is roughly constant, the heat
current that flows into the thermometer can be written as follows:  
\be
J_P^Q \propto \sum_{k=-1}^1 \int d\omega \phi_{lP}^Q(k,\omega) \left[
  f(\omega) - f_P(\omega_k)\right],
\ee
where
\be
\phi_l^Q(k,\omega) = (\omega_k - \mu) \varphi_{l}(k,\omega),
\ee
\be \label{varphi}
\varphi_{l}(k,\omega)  =   \sum_{\alpha= L,R}
 \left|G^R_{l, l \alpha}(k,\omega)\right|^2 \Gamma_{\alpha}(\omega).
\ee 

If the temperature $T$ of the reservoirs is small compared to their
Fermi energy, we can apply the Sommerfeld expansion up to order
$T^2$. 
Under this condition the heat current can be rewritten as
\ba \label{J2}
J_P^Q & \propto & \sum_{k=-1}^1 \left\{ \int_{\mu-k \Omega_0}^\mu d\omega
\phi_{lP}^Q(k,\omega) + \frac{\pi^2}{6} T^2 F_{lP}^Q(k,\mu)
\right. \nn \\   
& & - \left.\frac{\pi^2}{6} (T_{lP})^2 F_{lP}^Q(k,\mu-k\Omega_0)
\right\},
\ea
where
\be
F_{l}^Q(k,\omega) = \frac{d}{d\omega}\phi_l^Q(\omega).
\ee

The local temperature $T_{lP}$ corresponds to the solution of the
equation $J^Q_P (T_{lP}) = 0$. Directly from the expression for the
heat current given in Eq.(\ref{J2}) we can obtain $T_{lP}$: 
\be \label{localT}
(T_{lP})^2 \sim \frac{\frac{6}{\pi^2} \sum_k \Phi_{lP}(k) + T^2
  \sum_k F_{lP}^Q(k,\mu)}{\sum_k F_{lP}^Q(k,\mu-k\Omega_0)}, 
\ee
where
\be
\Phi_{l}(k) = \int_{\mu -k \Omega_0}^\mu d\omega \phi_{l}^Q(k,\omega). 
\ee

The adiabatic condition is introduced by expanding all the terms of
Eq. (\ref{localT}) in powers of the driving frequency $\Omega_0$.
It is easy to show that the first term of the numerator is of second
order in $\Omega_0$:
\be
\sum_{k=-1}^1 \Phi_{lP}(k) \approx \frac{1}{2} \Omega_0^2
\sum_{k=-1}^1 k\,\varphi_{lP}(k, \mu).
\ee

The second term of the numerator of Eq. (\ref{localT}) is
\be
\sum_{k=-1}^1 F_{lP}(k,\mu) = \sum_{k=-1}^1 \left[
  \varphi_{lP}(k,\mu) + k \Omega_0
  \frac{d\varphi_{lP}}{d\omega}(k,\mu) \right].
\ee

Expanding the denominator of Eq. (\ref{localT}) up to second order in
$\Omega_0$ we obtain
\ba
\sum_{k=-1}^1 F_{lP}^Q(k, \mu-k\Omega_0) & \approx & \sum_{k=-1}^1 \left[
  \varphi_{lP}(k,\mu) - k  \frac{d\varphi_{lP}}{d\omega}(k,\mu)
\Omega_0 \right.\nn \\ 
& & \left. + \frac{1}{2} k^2 \frac{d^2\varphi_{lP}}{d\omega^2}(k,\mu)
\Omega_0^2 \right].
\ea
Thus, keeping up to second order in $\Omega_0$ in Eq. (\ref{localT})
for the local temperature we obtain
\ba \label{localT2}
(T_{lP})^2 & \sim & \frac{3}{\pi^2} \lambda_{lP}^{(0)}(\mu)
\Omega_0^2 + T^2 \left( 1 + \right.\nn \\
& & \left. 2 \lambda_{lP}^{(1)}(\mu) \Omega_0 
- \frac{1}{2} \lambda_{lP}^{(2)}(\mu) \Omega_0^2\right),
\ea
where
\be
\lambda_{l}^{(n)}(\omega)=  \frac{1}{ \sum_{k=-1}^{1}
  \varphi_{l}(k,\omega)} 
\sum_{k=-1}^{1} (k)^{n+2} \frac{d^n[ \varphi_{l}(k,\omega)]}{d
  \omega^n} , 
\ee
and $\varphi_{l}(k,\omega)$ is given in Eq. (\ref{varphi}).

In particular, for the case of finite temperature $T$ of the
reservoirs, and high temperature compared to the driving ($T \gg
\Omega_0$), Eq. (\ref{localT2}) reduces to
\be \label{Tmu}
T_{lP} = T \left[ 1 + \lambda_{lP}^{(1)}(\mu) \Omega_0 \right].
\ee

For the case of reservoirs at very low temperature ($T \ll \Omega_0$),
Eq. (\ref{localT2}) leads to
\be \label{def1T0}
\Delta T_{lP} = \sqrt{\frac{3}{\pi^2} \lambda_{lP}^{(0)}(\mu)}
\Omega_0 - T. 
\ee

\subsection{Local temperature determined simultaneously with local
  chemical potential of the thermometer} \label{def2}

An alternative definition of local temperature to the one given in
Sec. \ref{def1} is the following: the local temperature $T_{lP}^*$
and the local chemical potential $\mu_{lP}^*$ are the values of the
temperature and the chemical potential of the probe that vanish
simultaneously $J^Q_P$ and $J^e_P$:
\be \label{currents0}
\left\{ \begin{array}{rcl}
J^Q_P (T_{lP}^*,\mu_{lP}^*) & = & 0, \\\nonumber
J^e_P (T_{lP}^*,\mu_{lP}^*) & = & 0.
\end{array} \right.
\ee

As we did in Sec. \ref{def1} we only keep terms up to order
$(V_0)^2$ for the weak-driving regime. 
Treating the coupling to the thermometer $w_{cP}$ at the lowest order
in perturbation theory and considering that the spectral function of
the thermometer $\Gamma_P(\omega)$ is roughly constant, the energy and
charge currents that flow into the thermometer can be written as
follows: 
\be \label{currents}
J_P^X \propto \sum_{k=-1}^1 \int d\omega \phi_{lP}^X(k,\omega) \left[
  f(\omega) - f_P(\omega_k)\right] ,
\ee
where $X = Q,e$ and 
\be
\phi_l^Q(k,\omega) =  (w_k - \mu_P) \varphi_{l}(k,\omega),
\ee
\be
\phi_l^e(k,\omega) =  \varphi_{l}(k,\omega),
\ee
where $\varphi_l(k,\omega)$ is given in Eq. (\ref{varphi}).

Applying the Sommerfeld expansion up to order $T^2$ and defining
$\mu_{lP}^* \equiv \mu + \Delta \mu_{lP}^*$, Eq. (\ref{currents}) can
be rewritten as
\ba \label{currents2}
J_P^X & \propto & \sum_{k=-1}^1 \left\{ \int_{\mu_-k\Omega_0 + \Delta
  \mu_{lP}^*}^\mu d\omega \phi_{lP}^X(k,\omega) + \frac{\pi^2}{6} T^2
F_{lP}^X(k,\mu) \right. \nn \\  
& & - \left. \frac{\pi^2}{6} (T_{lP}^*)^2
F_{lP}^X(k,\mu-k\Omega_0+\Delta \mu_{lP}^*) \right\} ,
\ea
where
\be
F_{l}^X(k,\omega) = \frac{d}{d\omega}\phi_l^X(\omega).
\ee

We expect $\Delta \mu_{lP}^*$ to be at least of order $\Omega_0$. We
expand the first term of Eq. (\ref{currents2}) up to second order in
$\Omega_0$:
\ba
\int_{\mu-k\Omega_0 + \Delta \mu_{lP}^*}^\mu d\omega
\phi_{lP}^X(k,\omega) & = &  - \phi_{lP}^X(k, \mu) (\Delta \mu_{lP}^*
- k \Omega_0) \nn \\ 
& & - \frac{1}{2} F_{lP}^X(k,\mu) (\Delta \mu_{lP}^* - k
\Omega_0)^2. \nn 
\\ 
\ea

In the case of finite temperature $T$ of the reservoirs, we define
$T_{lP}^* \equiv T + \Delta T_{lP}^*$ and expect $\Delta T_{lP}^*$ to
be at least of order $\Omega_0$. Hence, to the lowest order in
$\Omega_0$, $J^e$ and $J^Q$ become
\be \label{currents3}
\left\{ \begin{array}{rcl}
J^e_P (T_{lP}^*,\mu_{lP}^*) & \propto & - a \Delta \mu_{lP}^* - b
\Delta T_{lP}^* + \alpha \Omega_0, \\
J^Q_P (T_{lP}^*,\mu_{lP}^*) & \propto & - c \Delta \mu_{lP}^* - d
\Delta T_{lP}^* + \beta \Omega_0,
\end{array} \right.
\ee
where
\ba
a & = & \sum_{k=-1}^1 \left[ \varphi_{lP}(k,\mu) + \frac{\pi^2}{6}
  T^2 \frac{d^2 \varphi_{lP}}{d \omega^2}(k,\mu) \right], \\ 
b & = & \frac{\pi^2}{3} T \sum_{k=-1}^1
\frac{d\varphi_{lP}}{d\omega}(k,\mu), \\ 
c & = & T \sum_{k=-1}^1 \frac{d\varphi_{lP}}{d\omega}(k,\mu),
\\  
d & = & \sum_{k=-1}^1 \varphi_{lP}(k,\mu), \\
\alpha & = & \sum_{k=-1}^1 k \left[ \varphi_{lP}(k,\mu) +
  \frac{\pi^2}{6} T^2 \frac{d^2\varphi_{lP}}{d
    \omega^2}(k,\mu) \right],\\ 
\beta & = & T \sum_{k=-1}^1 k \frac{d\varphi}{d\omega}(k,\mu).
\ea

Within the approximations, the solution of the equations given in
Eq. (\ref{currents0}) is
\be 
\left\{ \begin{array}{rcl}
\Delta \mu_{lP}^* & = & \frac{1}{\Delta} (d \alpha - b \beta) \Omega_0,
\\ \\ 
\Delta T_{lP}^* & = & \frac{1}{\Delta} (a \beta - c \alpha) \Omega_0,
\end{array} \right.
\ee
where 
\be
\Delta = a d - b c = \left( \sum_k \varphi_{lP}(k,\mu) \right)^2 +
O(T)^2.
\ee

Hence, 
\be 
\left\{ \begin{array}{rcl}
\Delta \mu_{lP}^* & = & \left( \frac{\sum_k k \varphi_{lP}(k,\mu)}{\sum_k
  \varphi_{lP}(k,\mu)} + O(T)^2 \right) \Omega_0, \\ \\
\Delta T_{lP}^* & = & T \left[ \frac{d}{d\omega}\left(\frac{\sum_k k
    \varphi_{lP}(k,\omega)}{\sum_k \varphi_{lP}(k,\omega)}
  \right)_{\omega=\mu} + O(T)^2 \right] \Omega_0.
\end{array} \right.
\ee

Considering that $\left. \frac{d}{d\omega} \Gamma_\alpha
\right(\omega)|_{\omega = \mu} \sim 0$, then $T_{lP}^*$ becomes 
\be
T_{lP}^* = T \left[ 1 + \lambda^{(1)}_{lP}(\mu) \Omega_0 \right],
\ee
which coincides with Eq. (\ref{Tmu}).

For the case of $T \ll \Omega_0$, we propose the following ansatz for 
$\Delta \mu_{lP}^*$ and $\Delta T_{lP}^*$:
\be \label{ansatz}
\left\{ \begin{array}{rcl}
\Delta \mu_{lP}^* & = & \Delta \mu_0 + k_1 T,
\\ \\ 
\Delta T_{lP}^* & = & \Delta T_0 + k_2 T.
\end{array} \right.
\ee

We introduce in Eq. (\ref{currents2}) the values of $\Delta
\mu_{lP}^*$ and $\Delta T_{lP}^*$ given in Eq. (\ref{ansatz}). 
The result of this is expressions for the currents $J^e_\alpha$ and
$J^Q_\alpha$ in powers of $T$. 
Keeping terms up to first order in $T$ we can write the currents as
\ba
J^e_{\alpha} & = & J^{e,(0)}_{\alpha} + J^{e,(1)}_{\alpha}T, \\
J^Q_{\alpha} & = & J^{Q,(0)}_{\alpha} + J^{Q,(1)}_{\alpha} T.
\ea
The equations to be satisfied are four:
\be
J^{X,(n)} = 0,
\ee
where $X=e,Q$ and $n=0,1$.
The equations with $n=0$ lead to the values of $\Delta \mu_0$ and
$\Delta T_0$.
While the equations with $n=1$ lead to $k_1 = 0$ and $k_2 = -1$. 
Hence, $\Delta \mu_{lP}^*$ and $\Delta T_{lP}^*$ can be written as
\be \label{def2T0}
\left\{ \begin{array}{rcl}
\Delta \mu_{lP}^* & = & \frac{\sum_k k \varphi_{lP}(k,\mu)}{\sum_k
  \varphi_{lP}(k,\mu)} \Omega_0 ,
\\ \\ 
\Delta T_{lP}^* & = & \sqrt{\frac{3}{\pi^2} \lambda^{(0)}_{lP}(\mu)}
\Omega_0 - T. 
\end{array} \right.
\ee

The value obtained for $\Delta T_{lP}^*$ coincides with the one
obtained with the definition of local temperature given in
Sec. \ref{def1} [see Eq. (\ref{def1T0})].  



\begin{thebibliography}{99}
\bibitem{thermoqpoint}
L. W. Molenkamp, Th. Gravier, H. van Houten, O. J. A. Buijk,
M. A. A. Mabesoone, and C. T. Foxon, 
Phys. Rev. Lett. {\bf 68}, 3765 (1992); 
R. S\'anchez and M. B\"uttiker, 
Phys. Rev. B {\bf 83}, 085428 (2011)

\bibitem{liliheatpump}
L. Arrachea, M. Moskalets, and L. Martin-Moreno, 
Phys. Rev. B {\bf 75}, 245420 (2007).

\bibitem{heatquantumcap}
M. Moskalets and M. B\"uttiker,
Phys. Rev. B {\bf 80}, 081302 (2009). 

\bibitem{pekkola} 
F. Giazotto, T. T. Heikkil\"a, A. Luukanen, A. M. Savin, and
J. P. Pekola,   
Rev. Mod. Phys. {\bf 78}, 217 (2006); 
J. P. Pekola and F. W. J. Hekking,
Phys. Rev. Lett. {\bf 98}, 210604 (2007). 

\bibitem{dubi-diventra} 
Y. Dubi and M. Di Ventra, 
Nano Lett. {\bf 9}, 97 (2009);
e-print arXiv:0910.0425[cond-mat/] (to be published).

\bibitem{nanomec}
C. W. Chang, D. Okawa, A. Majumsar, and A. Zettl, 
Science {\bf 314}, 1121 (2006); 
A. Dhar, 
Adv. in Phys. {\bf 57}, 457 (2008);
D. Segal, 
Phys. Rev. Lett {\bf 101}, 260601 (2008); 
C. Chamon, E. Mucciolo, L. Arrachea and R. Capaz,
e-print arXiv:1006.4874v2[cond-mat/] 
(Phys. Rev. Lett., {\em in press}).

\bibitem{phot} 
T. Ojanen and A-P Jauho,
Phys. Rev. Lett. {\bf 100}, 155902 (2008). 

\bibitem{fdr} 
L. F. Cugliandolo and J. Kurchan, 
Phys. Rev. Lett. {\bf 71}, 173 (1993); 
Philos. Mag. B {\bf 71}, 501 (1995).

\bibitem{cukupel}
L. F. Cugliandolo, J. Kurchan, and L. Peliti, 
Phys. Rev E {\bf 55}, 3898 (1997); 
L. F. Cugliandolo and J. Kurchan, 
Physica A {\bf 263}, 242 (1999).

\bibitem{letoandco} 
H. Makse and J. Kurchan, 
Nature (London) {\bf 415}, 614 (2002); 
A. B. Kolton, R. Exartier, L. F. Cugliandolo, D. Dominguez, and
N. Gronbech-Jensen,  
Phys. Rev. Lett. {\bf 89}, 227001 (2002);
F. Zamponi, G. Ruocco, and L. Angelani, 
Phys. Rev. E {\bf 71}, 020101(R) (2005); 
L. Berthier and J.-L. Barrat, 
Phys. Rev. Lett. {\bf 89}, 95702 (2002);  
D. Segal, D. R. Reichman, and Andrew J. Millis, 
Phys. Rev. B {\bf 76}, 195316 (2007); 
R. A. Duine, 
{\em ibid.} {\bf 77}, 014409 (2008).
C. Aron, G. Biroli, and L. F. Cugliandolo, 
Phys. Rev. Lett. {\bf 102}, 050404 (2009).

\bibitem{letogus}
L. F. Cugliandolo and G. S. Lozano, 
Phys. Rev. Lett. {\bf 80}, 4979 (1998); 
Phys. Rev. B {\bf 59}, 915 (1999).

\bibitem{lilileto}
L. Arrachea and L. F. Cugliandolo, 
Europhys. Lett. {\bf 70}, 642 (2005).

\bibitem{cal} 
A. Caso, L. Arrachea, and G. S. Lozano, 
Phys. Rev. B {\bf 81}, 041301(R) (2010).

\bibitem{pump} 
L. J. Geerligs,  V. F. Anderegg, P. A. M. Holweg, J. E. Mooij,
H. Pothier, D. Esteve, C. Urbina, and M. H. Devoret,  
Phys. Rev. Lett. {\bf 64}, 2691 (1990). 
M. Switkes, C. M. Marcus, K. Campman, and A. C. Gossard,
Science {\bf 293}, 1905 (1999); 
S. K. Watson, R. M. Potok, C. M. Marcus, and V. Umansky, 
Phys. Rev. Lett. {\bf 91}, 258301 (2003).
M. D. Blumenthal, B. Kaestner, L. Li, S. Giblin, T. J. B. M. Hanssen,
M. Pepper, D. Anderson, G. Jones, and D. A. Ritchie,  
Nat. Phys.  {\bf 3}, 343 (2007).

\bibitem{qcapexp} 
J. Gabelli, G. F\`eve, J.-M. Berroir, B. Placais, A. Cavanna,
B. Etienne, Y. Jin, and D. C. Glattli, 
Science {\bf 313}, 499 (2006); 
G. F\`eve, A. Mah\' e, J.-M. Berroir, T. Kontos, B. Placais,
D. C. Glattli, A. Cavanna, B. Etienne, and Y. Jin, 
{\em ibid.} {\bf 316}, 1169 (2007).

\bibitem{fourpoint}
H. L. Engquist and P. W. Anderson, 
Phys. Rev. B {\bf 24}, 1151 (1981);
T. Gramespacher and M. B\"uttiker, 
{\em ibid.} {\bf 56}, 13026 (1997); 
L. Arrachea, C. Na\'on, and M. Salvay, 
{\em ibid.} {\bf 77}, 233105 (2008).

\bibitem{fourpointfed}
F. Foieri, L. Arrachea, and M. J. Sanchez 
Phys. Rev. B {\bf 79}, 085430 (2009); 
F.Foieri and L.Arrachea, 
{\em ibid.} {\bf 82}, 125434 (2010).

\bibitem{liliflo}
L. Arrachea, 
Phys. Rev. B {\bf 72}, 125349 (2005);
L. Arrachea, 
Phys Rev. B {\bf 75}, 035319 (2007);
L. Arrachea and M. Moskalets, 
Phys. Rev. B {\bf 74}, 245322 (2006).

\bibitem{polianski}
M. Polianski and M. B\"uttiker, 
arXiv:1001.2492.

\bibitem{adia} 
P.W. Brouwer, 
Phys. Rev. B {\bf 58},  R10135  (1998); 
M. Moskalets, and M. B\"uttiker, 
{\em ibid.} {\bf 66}, 035306 (2002).

\end{thebibliography}
\end{document}